\documentclass[aps,prl,floatfix,showpacs,preprintnumbers,amsymb,a4paper,superscriptaddress,twocolumn]{revtex4-1}
\usepackage{amsmath}
\usepackage{graphicx}
\usepackage{amssymb}
\usepackage{amsmath}
\usepackage{epstopdf}
\usepackage{float}
\usepackage{color}

\newcommand{\be}{\begin{equation}}
\newcommand{\ee}{\end{equation}}

\newcommand{\dd}[1]{\mathrm{d}{#1}}

\newcommand{\ddn}[2]{\mathrm{d^{#1}}{#2}}
\begin{document}

\title{Proposal for the creation and observation of a ghost trilobite chemical bond}
\author{Matthew T. Eiles*}

\affiliation{Department of Physics and Astronomy, 
Purdue University, 47907 West Lafayette, IN, USA}
\author{Zhengjia Tong}

\affiliation{Department of Physics and Astronomy, 
Purdue University, 47907 West Lafayette, IN, USA}
\author{Chris H. Greene}
%\thanks{}

\affiliation{Department of Physics and Astronomy, 
Purdue University, 47907 West Lafayette, IN, USA}
\affiliation{ Purdue Quantum Center, Purdue University, West Lafayette, Indiana 47907, USA.}
\date{\today }

\begin{abstract}The ``trilobite'' type of molecule, predicted in 2000 and observed experimentally in 2015, arises when a Rydberg electron exerts a weak attractive force on a neutral ground state atom. Such molecules have bond lengths exceeding 100 nm.  The ultra-long-range chemical bond between the two atoms is a nonperturbative linear combination of the many degenerate electronic states associated with high principal quantum numbers, and the resulting electron probability distribution closely resembles a fossil trilobite from antiquity. We show how to coherently engineer this same long-range orbital through a sequence of electric and magnetic field pulses even when the ground state atom is not present, and propose several methods to observe the resulting orbital. The existence of such a \textit{ghost} chemical bond in which an electron reaches out from one atom to a nonexistent second atom is a consequence of the high level degeneracy.  
\end{abstract}
\maketitle

Many stunning experiments in recent years have demonstrated that a novel type of chemical binding occurs between a highly excited Rydberg atom and a neutral ground-state atom \cite{Bendkowsky,BoothTrilobite,Butterfly}. The Fermi pseudopotential reveals that this weak bonding derives from the low-energy scattering of the Rydberg electron off the neutral atom, and furthermore describes this interaction with a delta function proportional to the $s$-wave electron-atom scattering length $a_s$ \cite{Fermi,Greene2000}. Rydberg molecules have been observed in Cs, Rb, and Sr, all of which have $a_s<0$ \cite{Bendkowsky,TallantCs,DeSalvo2015}. The most interesting Rydberg molecules are the highly polar varieties, dubbed ``trilobites'' and ``butterflies'' for the unusual appearance of their electronic densities \cite{HamiltonGreeneSadeghpour,KhuskivadzeJPB}. In these molecules degenerate non-penetrating high angular momentum $l$ states hybridize to maximize either the electron's probability (trilobite) or probability gradient (butterfly) near the ground state atom \cite{BoothTrilobite,Butterfly,PfauRaithel}.

The electronic wave function of one of these molecules having a bond length $R_b$ is
\be
\label{eq:trilobite}
\Psi( R_b,\vec r) =\mathcal{N}\sum_{l=l_0}^{n-1}c_{nl,x}^{b}\phi_{nl}(\vec r),
\ee
where $\mathcal{N}$ is a normalization constant, $l_0$ restricts the summation to degenerate states ($l_0 \approx 3$ in alkali atoms) and $\phi_{nl}(\vec r) = \frac{u_{nl}(r)}{r}Y_{l0}(\theta,\phi)$ are standard hydrogenic Rydberg wave functions. The label $x$ refers to the type of molecule (i.e. trilobite or butterfly). Due to cylindrical symmetry and the functional form of the pseudopotential, $m_l=0$. The coefficients $c_{nl,x}^b$ are determined by diagonalizing the Fermi pseudopotential in the basis of degenerate hydrogenic states. For example, $c_{nl,\text{trilobite}}^{b} = \phi_{nl}(\vec R_b)$. For a general target state these coefficients can be compactly expressed as a vector, $\vec c_T$. 

The degeneracy needed to form these exotic states is exact for all $l$ in nonrelativistic hydrogen. Since the hydrogen-electron scattering lengths are both positive, the repulsive trilobite potential curves cannot support vibrational bound states.  Nevertheless, theoretical evidence suggests that the Rydberg electron-atom interaction still evinces resonant behavior related to the stationary points of the potential curves \cite{EilesHyd,TaranaCurik}. These are located at $R_b$ satisfying $u_{n0}(R_b) = 0$ \cite{ChibisovPRL2}. The index $b$ thus labels a series of trilobite states with specific bond lengths and nodal structure; furthermore, at these $R_b$ the wave function is dominated by just a few eigenfunctions of the Schr\"{o}dinger equation in elliptical coordinates \cite{Granger}. Like trilobite molecules, butterfly molecules are bound by electron-atom scattering, although they additionally depend on a $p$-wave shape resonance in the electron-atom scattering and the form of the $p$-wave pseudopotential involves gradient operators acting on the wave function \cite{HamiltonGreeneSadeghpour,KhuskivadzeJPB,Butterfly}. 

This letter shows that it is possible to create these chemical bonding orbitals with the ground state atom absent, and for this reason we refer to the electronic wave function (equation \ref{eq:trilobite}) as a {\it ghost chemical bond}.  By employing a carefully engineered sequence of alternating magnetic and electric fields, we evolve the wave function from an isotropic $ns$ state into precisely the same orbital that would form a chemical bond if a ground state atom were located at $R_b$.  The time evolution is described via unitary operators in degenerate first-order perturbation theory. A gradient ascent algorithm derived from optimal control theory optimizes the field sequence to ensure excellent overlap with the target state.  Two detection mechanisms are proposed to image and study this chemical bond, either in the ``ghost'' or in the true trilobite molecule.  Atomic units are used throughout.

{\begin{figure}[t]
\includegraphics[scale = 0.4]{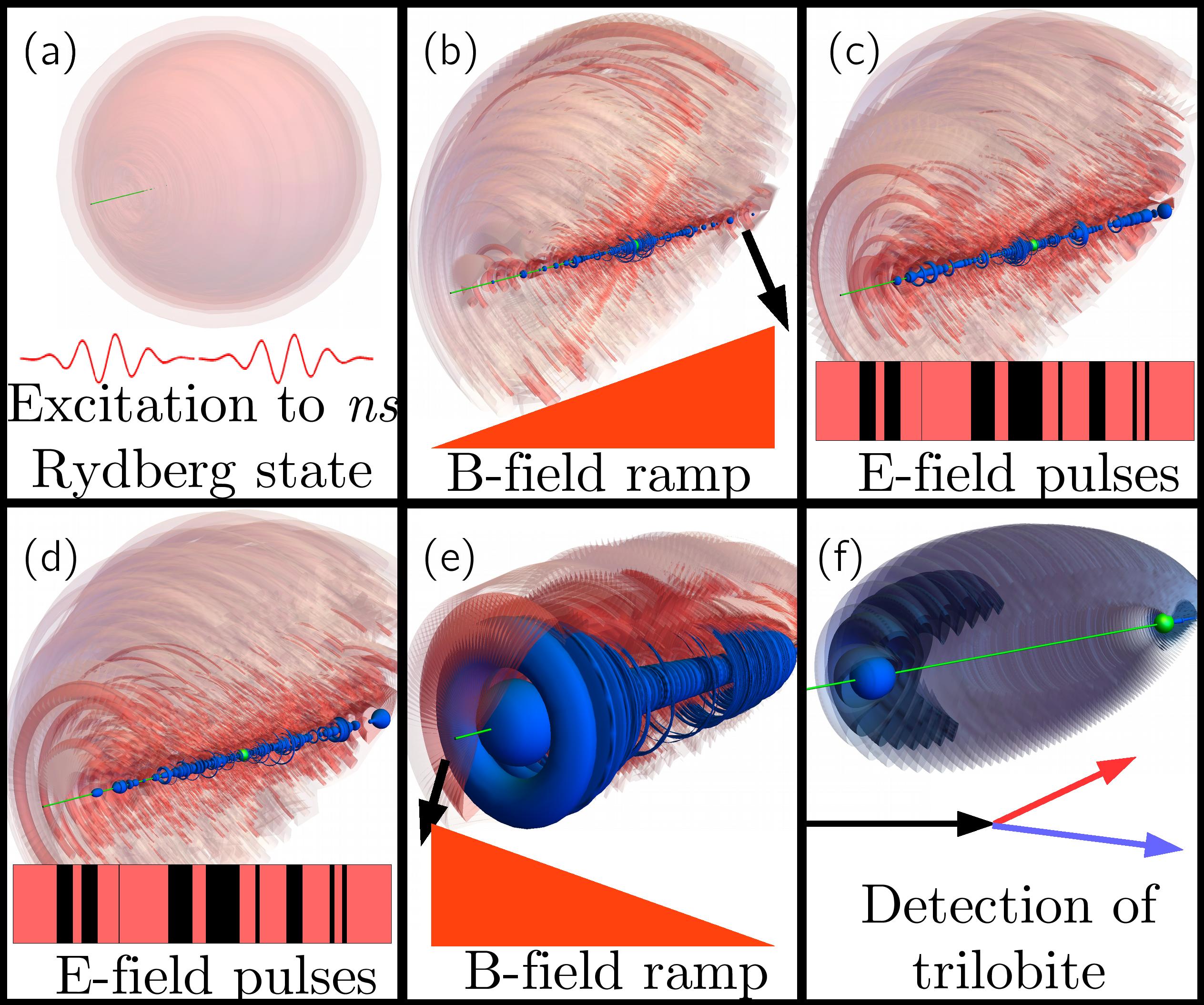}
\caption{\label{Fig:Scheme} The proposed scheme, illustrated using the $n=70,b=1$ trilobite as an example. The electronic probability is displayed in each panel usng three isosurfaces, defined where $|\Psi(x,y,z)|^2$ is $C$\% of $\text{Max}\left(|\Psi_\text{trilobite}(R_{b=1};x,y,z)|^2\right)$. They are cut away to reveal the inner structure. The Rydberg ion, not to scale, is the green sphere and the green line, parallel to the $z$ axis, provides a scale and is 1.1$\times 10^4a_0$ long. The color scheme for panels a-e is: bright blue when $C =1.54$, red when $C=0.154$, and translucent pink when $C = 1.54\times 10^{-2}$. For panel f: bright blue for $C = 15.4$, darker blue when $C=1.54$, and transparent blue when $C = 0.308$. a) First, an $ns$ Rydberg state is created.   b) Next, the magnetic field ramps on, creating a quadratic Zeeman state. c,d) After the magnetic field reaches its maximum value, many short electric field pulses are applied in addition to the magnetic field, creating complicated superpositions of the degenerate states. At different points in the sequence the wave function is strongly mixed. e) At the end of the sequence of electric field pulses, a proto-trilobite is created. The magnetic field ramps off and this state evolves into the trilobite state, f), which is detected. }
\end{figure}}

%%%
  Fig. \ref{Fig:Scheme} depicts the proposed experimental implementation of this concept. First, a hydrogen atom is excited to an $ns$ Rydberg state. 
Next, a magnetic field ramps on to a final amplitude $B$. Immediately after the ramp, a sequence of $N$  electric field pulses of amplitude $F$ are applied.  After the $N^{\text{th}}$ pulse the magnetic field ramps off. For the $n=70$ Rydberg state considered here, the ramp times are typically tens of $\mu$s, while the electric field durations and separations are several nanoseconds each, lasting in total several tens of $\mu$s as well. This whole process thus occurs within the natural radiative and blackbody lifetime of the Rydberg state. The trilobite state is particularly long-lived since it is an admixture of predominantly high$-l$ states and its decay rate is therefore mostly affected by blackbody radiation. We set a conservative lower lifetime bound at 200$\mu$s.  This time increases with decreasing ambient temperature, extending to several milliseconds at 10K  \cite{Lifetimes1,Blackbody1,Blackbody2,GallagherBook,RydbergBBRLifetime}.  As lower $l$ states bleed away, the dominant components of the trilobite state persist, leading to the remarkable scenario where the Rydberg electron remains localized in a small point several hundreds of nanometers away from the proton for many tens of microseconds. Dephasing caused by the small, MHz-scale energy splittings between fine structure levels of the low-$l$ states is limited since these components decay sooner, and additionally this dephasing is further eliminated since the final state is predominately high-$l$ states. These small energy splittings could impact the fidelity as these low-$l$ states are more prominent during the time evolution, but the presence of the large static magnetic field will prevent dephasing between different fine structure states due to the Paschen-Back effect \cite{Note1}. Interesting effects arise if a small electric field is pulsed on in this final state. The ghost chemical bond will revive every $\tau = \frac{2\pi}{3Fn}$, which is about 37 ns for a $0.1$V/cm electric field. Furthermore, the decay mode of the ghost molecule will change as the trilobite state is a linear combination of non-degenerate Stark eigenstates.  Details describing the physical considerations guiding this scheme design and some of the relevant parameters are found in the supplementary material \cite{Note1}.

Fig. \ref{Fig:orbitals} displays several ghost chemical bonds which can be created by this scheme. Perhaps the most distinctive characteristic of trilobite bonds is their nodal structure: as $b$ increments by one, a new lobe in the direction perpendicular to the intermolecular axis appears (Fig.\ref{Fig:orbitals}a shows a $b=3$ trilobite). Moreover, they are remarkably localized, maximally so in the $b=1$ trilobite shown in Fig. \ref{Fig:Scheme}f, where 20\% of the electron density occupies a region around the ghost atom smaller than 0.1\% of the total classically allowed volume. This localization is because the trilobite state, by construction, is the representation of the three-dimensional delta function in the finite basis of hydrogenic states in a single $n$-manifold.  The butterfly molecule chemical bond  (Fig. \ref{Fig:orbitals}b) has a bond length an order of magnitude smaller than the trilobite, and the wave function, fanning out into a winglike structure, fills much more of the classically allowed volume.  Rydberg molecules need not be exclusively diatomic: polyatomic Rydberg trilobite or butterfly molecules with more than one ground state atom lying within the Rydberg wave function have been studied theoretically \cite{Rost2006,JPBdens} but are challenging to observe in an experiment due to the low probability of finding the right configuration of three atoms. This restriction is of course lifted for bonds to nonexistent ghost atoms (Fig. \ref{Fig:orbitals}c) The coefficients $c_{nl,x}^b$ for these other Rydberg chemical bonds are provided in the supplementary material \cite{Note1}, and it should be remembered that the proposed method is generic to \textit{any} set of coefficients since the key requirement is that only the degenerate $n$-manifold is included. Time-dependent wave packets or the exotic giant dipole states, hydrogen atoms exposed to crossed fields such that the electron localizes in an anisotropic harmonic oscillator potential extremely far from the nucleus, are not allowed in this scheme as they are superpositions of non-degenerate states \cite{GiantDipole}. The Stark state (Fig. \ref{Fig:orbitals}d) and the Zeeman state (Fig. \ref{Fig:Scheme}b) highlight that even localized electron wave functions in static fields are entirely different from Rydberg molecule wave functions. 

\begin{figure}[b]
\includegraphics[scale = 0.3]{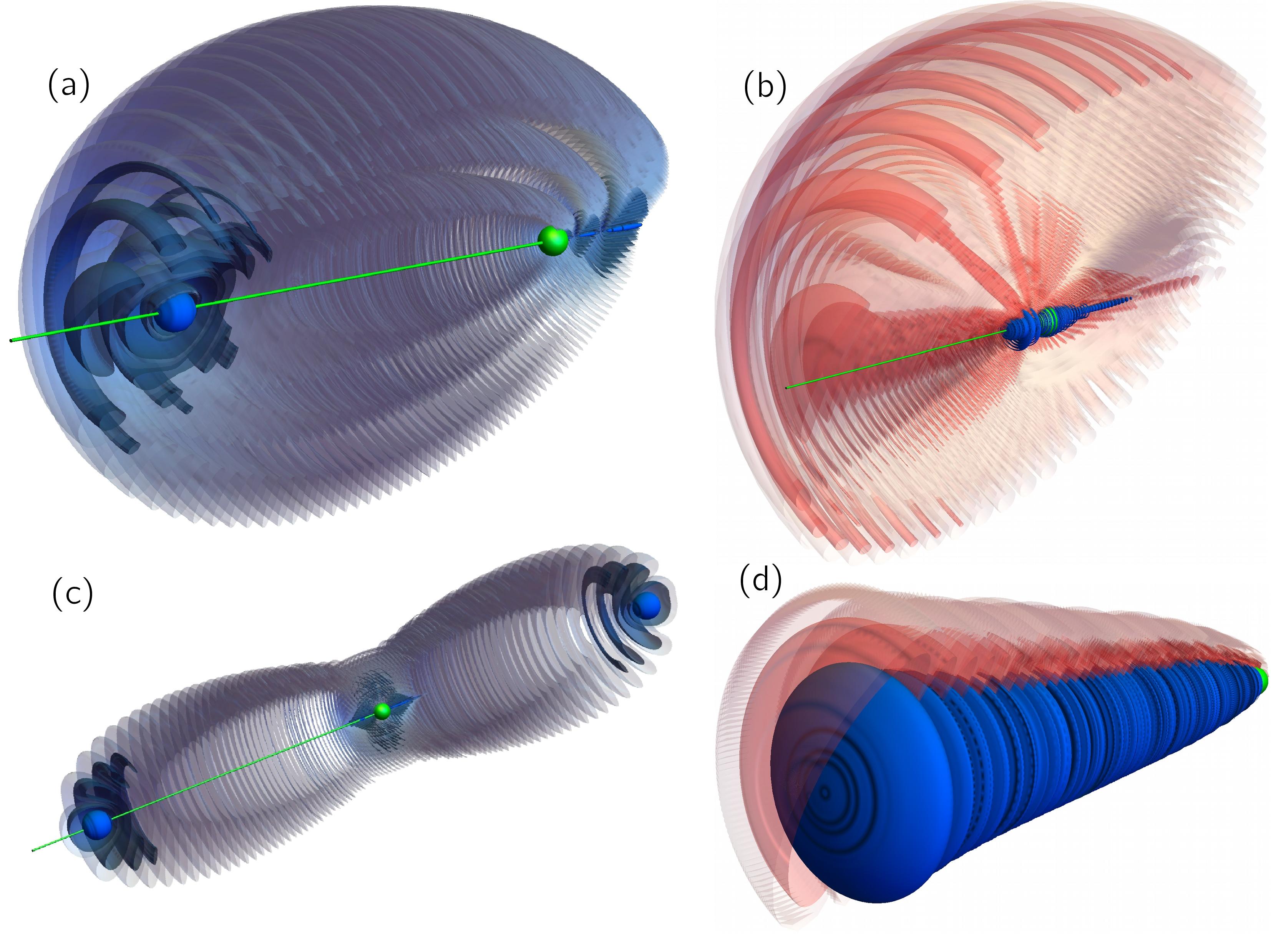}
\caption{\label{Fig:orbitals} A gallery of trilobite-like ghost bonds  for $n=70$, shown as isosurfaces as described in the caption of Fig. \ref{Fig:Scheme}. (a) A $b=3$ trilobite; (c) an even-parity collinear $b=1$ trilobite trimer.  (b) a butterfly with $R_0 = 653$; (d) the deepest Stark state.  
}
\end{figure}
The preparation of these exotic chemical bonds hinges on the fact that Rydberg electrons are strongly affected by external fields. These can manipulate the wave function into classical wave packets \cite{Stroud} or long-lived circular states \cite{Circular,CircNew}.  The enormous extent of Rydberg wave functions creates large transition dipole matrix elements, facilitating easy control even with weak field strengths. The Stark and quadratic Zeeman matrix elements scale as $Fn^2$ and $B^2n^4$, respectively, where $F$ and $B$ are electric and magnetic field amplitudes \cite{GallagherBook}. 

{\begin{figure}[t]
\includegraphics[scale = 0.31]{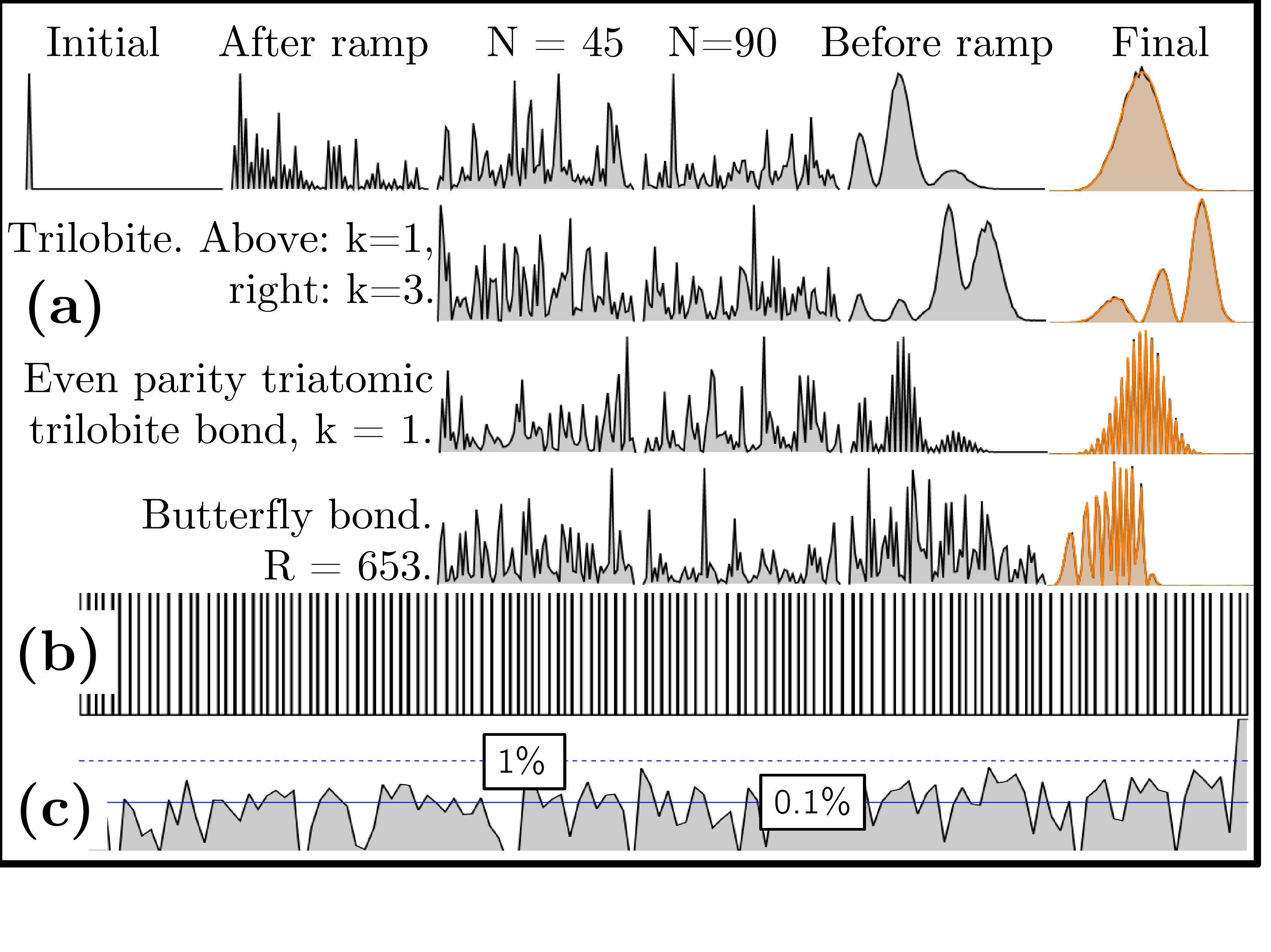}\\
\vspace{-20pt}
\caption{\label{Fig:schemedetails} Details of the proposed scheme for the four orbitals shown in Fig \ref{Fig:orbitals}. a) $l$-distributions, spanning from $l=0$ on the left to $l = 69$ on the right, at six different times. The first two times, the initial state and the Zeeman state following the field ramp, are identical in all cases. The orange overlay in the final state shows the exact target distribution.  b) The field pulses responsible for this $b=3$ trilobite. The electric fields are turned on in black regions and off in white regions. c) The fidelity, on a logarithmic scale, as a function of time.  }
\end{figure}}

The goal of our control scheme is to engineer a final state $\vec c_f$ which matches the target state, $\vec c_T$. Their similarity is characterized by the fidelity $\Phi = |\langle \vec c_T|\vec c_f\rangle|^2$.  After choosing the field amplitudes and initial ramp times the final state is determined by the $2N$ time periods: $\Delta t_i^f$, when both fields are on, and $\Delta t_i^b$, when only the magnetic field is on.  Although square pulses are used for simplicity in the present calculations, the error of a finite ramp time in an experiment should still be within the tolerance of our ideal parameter scheme, and a more thorough calculation could certainly include arbitrary pulse shapes to better match experimental conditions. A gradient ascent algorithm finds optimal $\Delta t_i^{b,f}$ parameters giving local maxima in $\Phi$ remarkably efficiently. Numerical experiments reveal several generic features of this approach. First, every optimal pulse sequence is primarily determined by the distribution of initial values, typically drawn from a uniform distribution of experimentally realistic values. This implies that there are effectively infinitely many good pulse configurations. One of these is shown in Fig. \ref{Fig:schemedetails}b, and a full data table of this sequence along with others which create the other chemical bonds in Fig. \ref{Fig:orbitals} is given in the supplementary material \cite{Note1}.  The non-uniqueness of the solutions implies that $\Phi$ is not convex, so there is no guarantee that a gradient ascent algorithm will discover global maxima. Surprisingly, our simulations found that, provided enough pulses (typically $N \approx 2n$) of adequate duration ($\sim 10$s of ns) are used, almost all sequences gave $\Phi > 0.999$.  Finally, methods that attempt to increase the fidelity monotonically with each pulse appear impossible. Fig. \ref{Fig:schemedetails} c shows $\log_{10}\Phi$: at no time prior to the final step of evolution does the fidelity increase above 0.01, nor is it monotonically increasing.

These findings are corroborated by optimal control theory \cite{BrifRabitz}. Quite general quantum proofs exist demonstrating that the topology of quantum control landscapes are very favorable to simple extrema search algorithms  \cite{RabitzInf,optimcontrol}. Only globally maximal seams exist in a sufficiently large parameter space; local maxima do not exist \cite{RabitzQuantumControl,Rabitz1,OCTstuff}. It is justifiable to restrict the initial guess to realistic experimental values for the field strengths and durations, rather than directly implementing these constraints into the search algorithm. This works excellently given the incredible flexibility of possible solutions \cite{RabitzInf,RabitzJCP}. The lack of local maxima guarantees that optimal solutions are found rapidly without more complicated genetic algorithms or stochastic optimizers \cite{RabitzQuantumControl}.

The complex quantum pathways the wave function evolves along necessitate very stringent control over the field amplitudes and pulse timing. The experiment must be very well shielded from stray fields so that the control field amplitudes can be specified to better than 10 $\mu$V/cm and 1 mG. Rydberg atoms themselves can be used as highly sensitive field sensors \cite{HarocheE,BfieldRydRaithel}. The pulse timing should be controlled to femtosecond precision. These error bounds correspond to a 10\% reduction in the fidelity from the theoretical prediction. This sensitivity appears to be intrinsic to high Rydberg states, rather than caused by a poorly informed optimal control theory approach. These high sensitivities may require more precise theoretical methods to compute the time evolution as effects beyond first-order perturbation theory are near to this level of accuracy; the proof of principle demonstrated here is still applicable in more sophisticated approaches. A straightforward improvement could include a a robustness measure as a cost function in the optimization, and then use more sophisticated optimization techniques.  Recently this approach successfully obtained optimal radio-frequency pulses to excite circular Rydberg states \cite{CPKoch2018}, and could similarly ease the experimental difficulties here.

\begin{figure}[t]
\includegraphics[scale = 0.45]{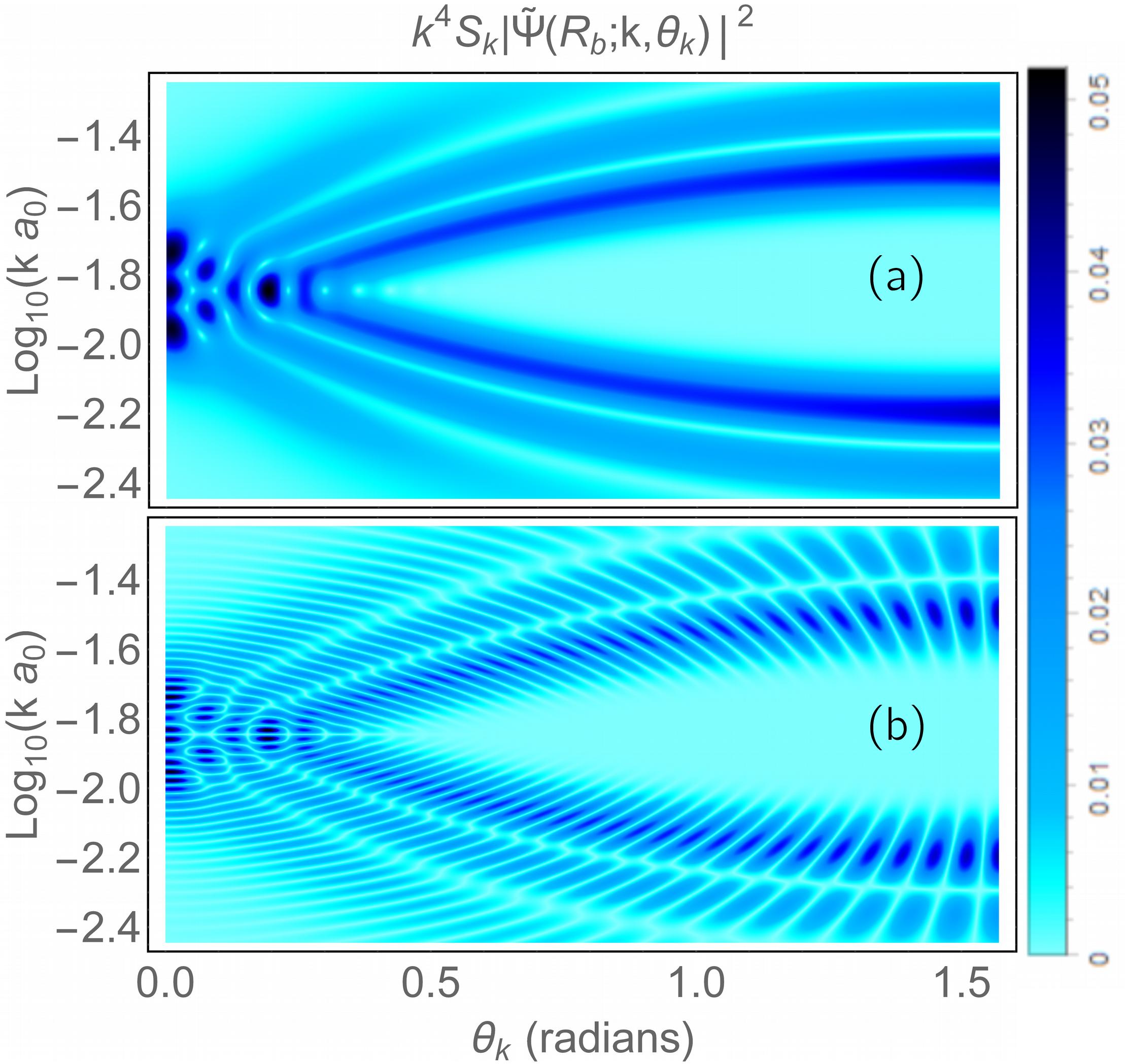}
\caption{\label{Fig:MomSpaceTwo}Momentum-space probability distributions for the $n=70$  $b=3$ trilobite dimer (a) and trimer (b).  Both are symmetric about $\theta_k=\frac{\pi}{2}$ and, when multiplied by $k^4$, are logarithmically symmetric about $k=1/n$. The scaling factor $S_k=(\theta_k + 0.1)$ enhances the visibility at large $\theta_k$. 
}
\end{figure}

Two experimental methods could directly detect the ghost orbital: electron momentum $(e,2e)$ spectroscopy and $x$-ray diffraction \cite{Note1}. In $(e,2e)$ spectroscopy an energetic electron scatters from and ejects the Rydberg electron; both are detected in coincidence \cite{e2epaper1,e2ereview1,ChemPhyse2e}. If the electron-electron collision is fully elastic--requiring large momenta, energies exceeding the ionization potential, and large momentum transfer between the electrons--the triply differential cross section is proportional to the electron's momentum density \cite{e2epaper1,ChemPhyse2e}.  Typically only spherically averaged cross sections can be measured in isotropic samples, but since these trilobite-like orbitals are aligned in the direction of the control fields the fully differential cross section can be measured.

\begin{figure}[h]
\includegraphics[scale = 0.45]{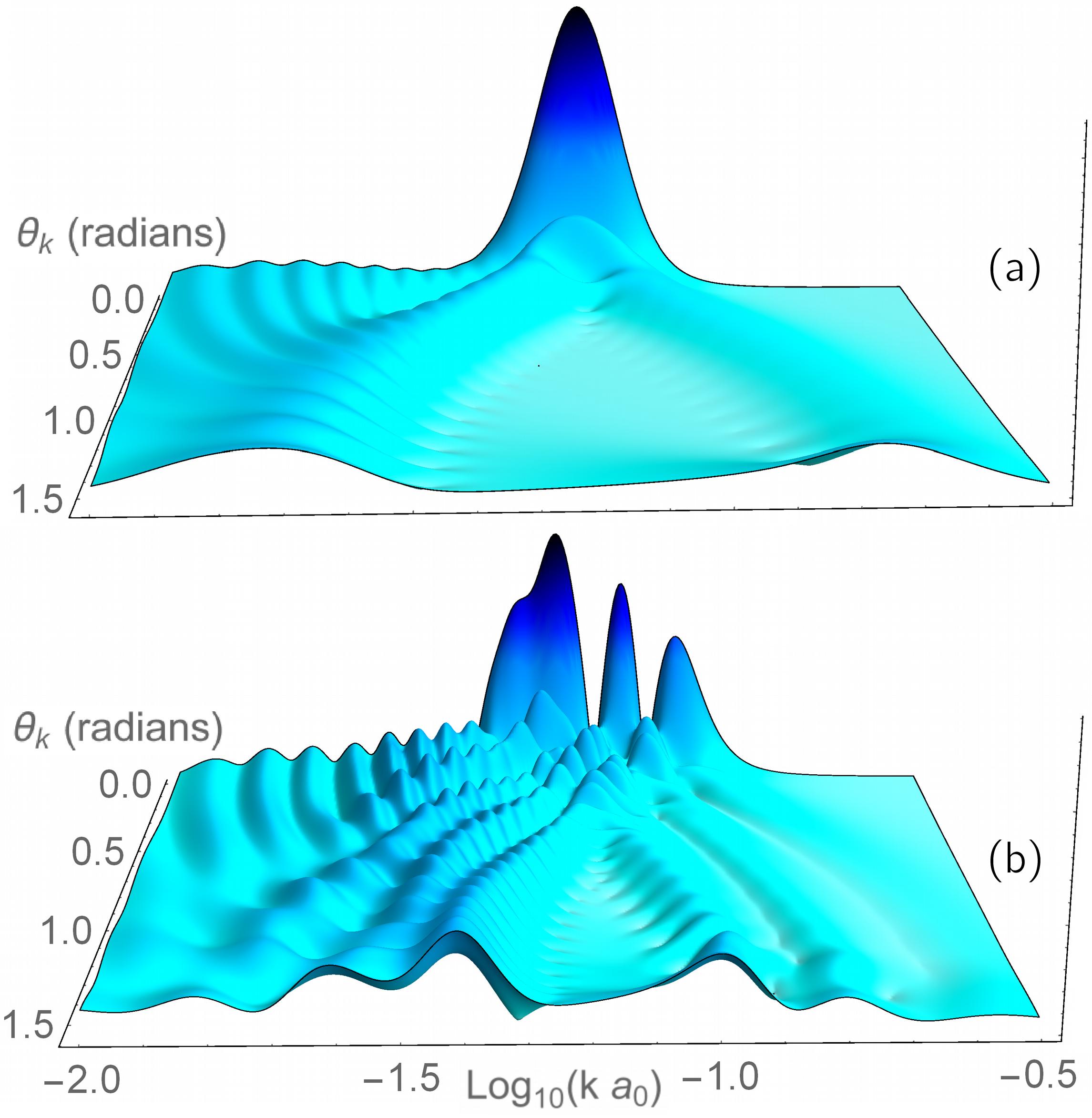}

\caption{\label{Fig:MST} The quantity $\left|k^2F_T\left(R_b;\vec k\right)\right|$ for the $n=30$  $b=1$ and $b=3$ trilobites. The scaling factor $k^2$ is added to improve visibility of small features. % of (a) is $2\times 10^{-4}$, and (b) is $1.5\times 10^{-4}$
This function is also symmetric about $\theta_k=\frac{\pi}{2}$. $n=30$ was chosen to connect back to the very first trilobite molecule prediction \cite{Greene2000}. 
}
\end{figure}
A complementary technique is $x$-ray diffraction \cite{xray}. The differential scattering cross section for this process is proportional to the Fourier transform of the electron density itself, creating another window into the electronic structure of these ghost orbitals. Figs. \ref{Fig:MomSpaceTwo} and \ref{Fig:MST} show these two different quantities--the momentum density and the Fourier-transformed position density, respectively--for several example ghost orbitals. As expected from Fourier analysis, the momentum-space wave functions in Fig. \ref{Fig:MomSpaceTwo} mirror the symmetries and nodal structure present in the real-space wave functions. Three ridges mirror the nodes in the $b=3$ trilobite, and the even-parity trimer possesses additional nodes overlapping these ridges showing the absence of odd-parity components. The symmetries in $\theta_k$ and $k$ relate to the symmetry of the real-space wave function. The Fourier-transformed electron density has many of these same features, although it is no longer symmetric about $k=1/n$. It is also significantly smaller in magnitude and less clear to interpret, although the major nodal features corresponding to the trilobite lobes are still apparent \cite{Note1}.

This letter discussed how exotic ultra-long-range ghost chemical bonds consisting of just one atom could be formed and detected in the laboratory. The electron can be either forced to localize very tightly on one or more positions in space, as in the ``trilobite'', or to spread out into an exotic fan-like structure, as in the ``butterfly''.  The control scheme, consisting of a slowly varying magnetic field along with a sequence of rapidly pulsed electric fields, emulates the Fermi pseudopotential responsible for the formation of Rydberg molecules by neutral perturbers.  The specifics of the field timings are designed efficiently using a gradient ascent algorithm, and excellent fidelity ($\gg 99.9\%$) can be reached with typical laboratory fields and time scales. The stringent control constraints on the field amplitudes and pulse timings, requiring excellent shielding towards stray fields and knowledge of field strengths to high accuracy, are certainly major experimental hurdles reflecting the exaggerated energy scales and complexity of this system, a novel regime for optimal control theory. One can envision even more exotic ghost states for future study, such as extended configurations like the trimer molecules shown here with several more ``ghost'' atoms spaced along a line \cite{Note1}, or even non-cylindrical polyatomic orbitals (requiring $m_l\ne 0$ contributions also).``Electron lattices'' could even be envisioned where wave function peaks are placed regularly around the atomic core, with potential applications in quantum gate technology. The theory could be likewise extended to atoms with quantum defects or performed more accurately to include nonperturbative field effects. The proposed detection methods are equally applicable to real Rydberg molecules \cite{Hua1, Dixit}.

\acknowledgements
We are pleased to acknowledge enlightening discussions with F. Robicheaux, R.T. Sutherland, and F. Jafarpour. This work was supported in part by NSF Grant No.PHY-1607180.

%\bibliography{../../../../Prelim_Documents/Thesis_bib}

%

%%%%%%%%%% Merge with supplemental materials %%%%%%%%%%
\pagebreak
\widetext
\begin{center}
\textbf{\large Supplemental Information}
\end{center}
%%%%%%%%%% Merge with supplemental materials %%%%%%%%%%
%%%%%%%%%% Prefix a "S" to all equations, figures, tables and reset the counter %%%%%%%%%%
\setcounter{equation}{0}
\setcounter{figure}{0}
\setcounter{table}{0}
\setcounter{page}{1}
\makeatletter
\renewcommand{\theequation}{S\arabic{equation}}
\renewcommand{\thefigure}{S\arabic{figure}}
\renewcommand{\bibnumfmt}[1]{[S#1]}
\renewcommand{\citenumfont}[1]{S#1}

\section{introduction}
This supplementary material provides more information about the evolution of the wave function, the detection schemes, and the relevant physical parameters for a realistic implementation. It also provides the actual pulse durations, $\Delta t_i^f$ and $\Delta t_i^b$, used for the figures in the main text. Additional figures showing the correspondence between position and momentum representations of the wave functions are also given.

\section{Time Evolution}
We describe the time evolution caused by these time-dependent field pulses using degenerate first-order perturbation theory. The electronic wave function is expanded into the degenerate stationary Rydberg states of a given $n$ with time-dependent coefficients: $\Psi(\vec r,t) = \sum_lc_l(t)\phi_{nl}(\vec r)$.  In a parallel field configuration $m_l$ is still a good quantum number, so we consider only the $m_l = 0$ subspace. In a time period $\Delta t_i^b$ when the electric field is zero, the Hamiltonian is $H_{B} = H_0 + \frac{B^2}{8}r^2\sin^2\theta$. Likewise, when the electric field is nonzero for a time period $\Delta t_i^f$, $H_F =H_B + Fr\cos\theta$. In the degenerate subspace of a single $n$ manifold, the action of $H_0$ is irrelevant and can be set to zero; the operators $H_B$ and $H_F$ are then diagonalized to obtain the diagonal eigenvalue matrices $\underline{b}$, $\underline{f}$ and eigenvector matrices $\underline{U_b}$, $\underline{U_f}$, respectively. The integrals involved in finding the matrix elements of $H_B$ and $H_F$ are 
\begin{align}
\langle l|\cos\theta|l'\rangle &= \sqrt{\frac{(l_<+1)^2}{(2l_<+1)(2l_<+3)}}\delta_{l,l'\pm1},\\
\langle l|\sin^2\theta|l'\rangle &= \frac{2(l^2+l-1)}{(2l+3)(2l-1)}\delta_{ll'}-\sqrt{\frac{(l_<+2)^2(l_<+1)^2}{(2l_<+5)(2l_<+3)^2(2l_<+1)}}\delta_{l,l'\pm 2}\nonumber,\\
\langle nl|r^k|n'l'\rangle &= \frac{2^{l+l'+2}}{n^{l+2}n'^{l'+2}}\sqrt{\frac{(n-l-1)!(n'-l'-1)!}{(n+l)!(n'+l')!}}\\&\times\sum_{m,m'=0}^{m=n-l-1}\begin{pmatrix}n+l\\n-l-1-m\end{pmatrix}\begin{pmatrix}n'+l'\\n'-l'-1-m'\end{pmatrix}\frac{(k+m+m'+l+l'+2)!}{\left(\frac{n+n'}{nn'}\right)^{k+3+l+l'+m+m'}},
\end{align}
where $()$ are binomial coefficients. A derivation of the radial matrix element can found in \cite{radmatel}. 

During each $\Delta t$ the Hamiltonian is time-independent, so the evolution of the initial state is computed by iteratively acting on it with the unitary time evolution operator for each pulse until the final time is reached:
\begin{align}
\label{eq:pulses}
\vec {c_f} &= \underline{X_B}^{\dagger}\left[\prod_{i = 1}^N\underline{U_b}e^{-i\Delta t_i^b\underline{b}}\underline{U_b}^{-1}\underline{U_f}e^{-i\Delta t_i^f\underline{f}}\underline{U_f}^{-1}\right]\underline{X_B}\vec {c_0}.
\end{align}
$X_b = \underline{U_b}e^{-i\Delta t_\text{ramp}\underline{b}}\underline{U_b}^{-1}$ is the field ramp operator. 

 Unlike many problems where major effort is needed to numerically calculate the gradient \cite{GrapeAlgo}, in the present case the gradient is given analytically (without loss of generality here is the partial derivative with respect to $\Delta t_j^f$):
\begin{align}
\frac{\partial \Phi}{\partial \Delta t_j^f}&=\left\langle\frac{\partial \vec{c_f}}{\partial \Delta t_j^f}\right|\vec{c_T}\Bigg\rangle\langle \vec{c_T}|\vec{c_f}\rangle + \text{c.c.}\\
\frac{\partial \vec{c_f}}{\partial\Delta t_j^f} %&= \frac{\partial}{\partial \tilde{F_j}} \left[\prod_{i = 1}^N\underline{U_b}e^{-i\tilde{B_i}\underline{b}}\underline{U_b}^{-1}\underline{U_f}e^{-i\tilde{F_i}\underline{f}}\underline{U_f}^{-1}\right]\vec c_0\\
&= \underline{X_B}^{\dagger}\Bigg[\prod_{i = 1}^N\underline{U_b}e^{-i\Delta t_i^b\underline{b}}\underline{U_b}^{-1}\underline{U_f}\left(-i\underline{f}\right)^{\delta_{ij}}e^{-i\Delta t_i^f\underline{f}}\underline{U_f}^{-1}\Bigg]\underline{X_B}\vec {c_0}.
\end{align}
After finding this gradient, the full set of parameters $\{\Delta t_i^O\}$ is shifted in the direction of the steepest change in the fidelity, $\Delta t_i^O\to \Delta t_i^O + \varepsilon\frac{\partial{\Phi}}{{\partial \Delta t_i^O}}$, where $\varepsilon$ is a variable stepsize and $O$ represents either $f$ or $b$.

\section{Physical considerations for the control scheme}
\noindent\textit{Relativistic effects}\\ The fine structure breaks the exact degeneracy of $nl$ states with low $l$. The $70p_{1/2,3/2}$ ($70d_{3/2,5/2}$)  levels are split by $\approx 0.25(0.085)$ MHz. Precession between different $m_l$ states caused by the $np$ splitting over the time scale of the experiment introduces undesirable decoherence. To avoid this problem a magnetic field of 100G is applied to access the Paschen-Back regime and eliminate this precession by energetically separating $m_l\ne 0$ levels.
\\
\textit{Magnetic fields}\\ These relatively large magnetic fields are also necessary since the quadratic Zeeman term is very small compared to the linear Stark shift. It is challenging experimentally to change fields of this strength quickly due to the self-inductance of the electronics. We have adopted the slew rate 10G/5$\mu$s of state of the art magnetic field coils developed to quench ultracold atomic species \cite{MakotynThesis}.
\\
\textit{Pulse sequence and field strength parameters}\\ There is considerable flexibility in the initially guessed pulse distribution. The simplest theoretical choice is simply to draw the initial distribution randomly from a region of experimentally realistic values. The range of these values should be limited to reduce the overall time. The optimized field pulses then closely mirror this initial distribution due to the incredible multiplicity of optimal configurations and their sensitivity. Sequences leading to very good fidelity can also be found with an initial configuration of equal-duration pulses, although with somewhat more difficulty since the parameter space is less robust. Setting $N = 130$ essentially guaranteed $\Phi>0.999$ regardless of the initial distribution, while for $N = 120$ high fidelities $\Phi \approx 0.99$ were always achieved but $\Phi > 0.999$ was not. For $N$ even as low as 70 $\Phi \approx 0.9$ reliably. Likewise the field amplitudes can be adjusted to better conform to a specific experiment. The magnetic field amplitude should be fairly large, as described above, while the electric field amplitude should not be large enough that different Rydberg levels together can mix together. For $n = 70$ this restricts it to not far exceed the 0.1V/cm employed here. 
\\
\textit{Rydberg levels}\\ High principal quantum numbers are needed to ensure that the above constraints are sufficient, but also increase the theoretical complication and experimental limits on sensitivity. The benefits are contingent on various Rydberg scaling laws: the fine structure splitting decreases as $\mathcal{O}(n^{-3})$, magnetic field effects increase as $\mathcal{O}(n^4)$, and the natural and blackbody radiative decay rates decrease as $\mathcal{O}(n^{-3})-\mathcal{O}(n^{-5})$ (depending on $l$) and $\mathcal{O}(n^{-2})$, respectively. The number of pulses and electric field influence increase as $\mathcal{O}(n)$ and $\mathcal{O}(n^2)$, respectively, and the electron's momentum decreases as $\mathcal{O}(n^{-1})$; the sensitivity to error and difficulty of detection are thus increased at high $n$. Additionally, first-order perturbation theory starts to become less accurate at higher $n$ due to the decreasing (as $\mathcal{O}(n^{-3})$) separation between Rydberg manifolds. $n = 70$ functions in our exemplary calculations as a convenient middle ground, but the wide range of theoretical and experimental constraints mean that this choice is quite flexible depending on the circumstances.
\\
\textit{Calculation details}\\
 The specific field configurations for the orbitals in Fig. 1 in the main text are presented in table 1. We chose $n=70$, $B_{max}=100$G, $F_{max}=0.1$V/cm, and $N=130$.  $\Delta t_i^b$ and $\Delta t_i^f$ were chosen randomly from the ranges $\{200,400\}$ and $\{20,60 \}$ns, respectively.  The ability to of the gradient ascent algorithm within this scheme to find optimal solutions is surprisingly robust to variations of all these parameters. Scaling laws can also be used to vary the parameters. The system is invariant under the transformations $F\to \mathcal{F}$, $B\to B\sqrt{\mathcal{F}/F}$, $\Delta t_i^O\to \Delta t_i^O(F/\mathcal{F})$, and $T_\text{ramp} \to T_\text{ramp}(F/\mathcal{F})^{3/2}$. 
 \\
 \textit{Orbital details}\\
  The (unnormalized) final state coefficients are given by:\cite{JPBdens, ChibisovPRL2,KhuskivadzePRA}.
\begin{align}
\label{coefficients}
c_{nl,trilobite}^b&= \phi_{nl0}(R_b)\\
c_{nl,butterfly}^b&= \frac{\partial\phi_{nl0}}{\partial R_b}\\
c_{nl,even/odd parity trimer}^b&= \frac{(1\pm(-1)^l)}{\sqrt{2}}c_{nl,x}^b
\end{align}
evaluated at the bond positions $R_b$ which are determined for each state by finding the minima of $\sum_l|c_{nl,x}^b|^2$. 

The momentum-space wave function for a trilobite-like orbital characterized by coefficients $c_{nl,x}^b$ is
\begin{align*}
\tilde{\Psi}(R_b;\vec k) &= \frac{\sum_{l=l_0}^{l=n-1}c_{nl,x}^bF_{nl}(k)Y_{l0}(\theta_k,\phi_k)}{\left(\sum_{l=l_0}^{n-1} |c_{nl,x}^b|^2\right)};\\
F_{nl}(k) &= -(-i)^l\sqrt{\frac{2(n-l-1)!}{\pi(n+l)!}}n^22^{2(l+1)}l!n^lk^l\\&\times (n^2k^2+1)^{-l-2}C_{n-l-1}^{l+1}\left(\frac{n^2k^2-1}{n^2k^2+1}\right),
\end{align*}
where $C_i^j$ is the Gegenbauer polynomial of degree $i$ and order $j$; these ``radial'' momentum space wave functions were calculated shortly after the first solution of the quantum mechanical hydrogen atom \cite{PodolskyPauling}. They 
possess a symmetry under the change of variables $nk = a^x$, where $a$ is some constant. For simplicity, if $a$ is $e$,
\begin{align}
F_{nl}(x) = -(-i)^l\sqrt{\frac{2(n-l-1)!}{\pi(n+l)!}}n^22^{2(l+1)}l!n^lk^l e^{-2x}(\sinh(x))^{-l-2}C_{n-l-1}^{l+1}\left(\tanh(x)\right),
\end{align}
From inspection this function possesses a mirror symmetry (odd or even depending on the parity of $n$) about $x = 0$ when multiplied by $e^{2x}$. This is the reason for the symmetry about $k = 1/n$ of the function $|k^2\Psi(k)|^2$. 

\section{Further details about detection methods and momentum-space densities}
There are two additional detection techniques, involving field ionization, that could be employed instead of the two discussed in the main text. In Stark photoionization spectroscopy, the electron is photoionized and detected on a distant screen. This maps the real-space electron wave function into Stark states, but unfortunately the theoretical description and extraction of the wave function's properties is too imposing for this to be a robust and clear method in the present context \cite{PIspec,BordasFabrikant,Panos1}. A related, but more straightforward, technique is ionization by half-cycle pulses \cite{FrancisHCP,JonesHCP}. A half-cycle pulse along the $z$ axis imparts a momentum kick to the Rydberg electron, and if the electron's increased energy is sufficient to overcome the ionization potential it will be detected. A measurement of the electron current is therefore proportional to $\int_{k_z}^\infty \int_{k_x,k_y}|\Psi(\vec k)|^2\ddn{3}{k}$, where $\vec k$ is the momentum of the electron. A very similar observable is obtained in Compton scattering \cite{CooperCompton}. $k_z$ is determined by the momentum kick, so by varying this the probability density that the electron had momentum $k_z$ and any value of $k_x$ and $k_y$ can be obtained. Although the simplicity of this approach is attractive, much of the detailed structure of the wave function is averaged over and obscured. 

The advantage of both techniques we focused on in the main text is that they, in principle, can lead to direct measurements of the electron's spatial properties, specifically either its momentum density or the Fourier transform of its real-space density. Here we provide additional details of these two techniques. In $(e,2e)$ spectroscopy, an electron with momentum $\vec k_0$ collides with the Rydberg electron, which has momentum $\vec k$. Both elastically scatter into plane waves with momenta $\vec k_a$ and $\vec k_b$, and are detected in coincidence. The triply-differential cross section for this process is \cite{e2epaper1,e2ereview1,ChemPhyse2e}
\be
\label{eq:e2ecs}
\frac{d^3\sigma_{(e,2e)}}{d\Omega_a\Omega_bdE_b}= \frac{4k_ak_b}{k_0|\vec k_0-\vec k_a|^4}\left|\tilde{\Psi}(R_b;\vec k)\right|^2,
\ee 
where $\Omega_i$ is the solid angle in which electron $i$ is detected, and $E_b$ is the energy of electron $b$. There are many commonly used experimental geometries which determine the specific form of the kinematic factor $|\vec k_0 - \vec k_a|^{-4}$ and the relationship between $\vec k$, $\vec k_a$, and $\vec k_b$. Depending on the implementation, the angle of incident electrons or their initial energy can be tuned to vary $\vec k$. We have focussed on the key features of the momentum density, since this gives the properties of the Rydberg electron that we want to study, and the kinematic properties can then be optimized by the preferred choice of geometry. 

The differential scattering probability for $x$-ray diffraction is given by \cite{xray}
\be
\frac{d\sigma}{d\Omega} = \frac{d\sigma_{th}}{d\Omega}\left|\int\left|\Psi(R_b,\vec r)\right|^2e^{i\vec k\cdot\vec r}\ddn{3}{r}\right|^2,
\ee
where $\frac{d\sigma_{th}}{d\Omega}$ is the Thomson cross section. 
The Fourier transform of the electronic density is more involved than the calculation of the momentum wave functions. We proceed in the usual fashion, expanding the plane wave into spherical harmonics:
\begin{align*}
\int\ddn{3}{r}e^{i\vec k\cdot\vec r}|\Psi(R_k,\vec r)|^2 &= \sum_{l,k,p}\int\ddn{3}{r}c_lc_k4\pi i^pj_p(kr)u_{nl}(r)u_{nk}(r)Y_{l0}(\hat r)Y_{k0}^*(\hat r)Y_{p0}^*(\hat r)Y_{p0}(\hat k)\\
&= 4\pi \sum_{l,k}c_lc_k\sum_{p = |l-k|}^{p=l+k}i^p\sqrt{\frac{(2l+1)(2p+1)(2k+1)}{4\pi}}\begin{pmatrix}l & p & k\\ 0 & 0 & 0\end{pmatrix}^2\int\dd{r}j_p(kr)u_{nl}(r)u_{nk}(r)Y_{p0}(\hat k).
\end{align*}
Here $c_l,c_k$ are the coefficients determining the orbital state. 
This radial integral can be evaluated in terms of hypergeometric functions following Eq. 40 of Ref. \cite{Hua1}.

\begin{figure}
\includegraphics[scale = 0.48]{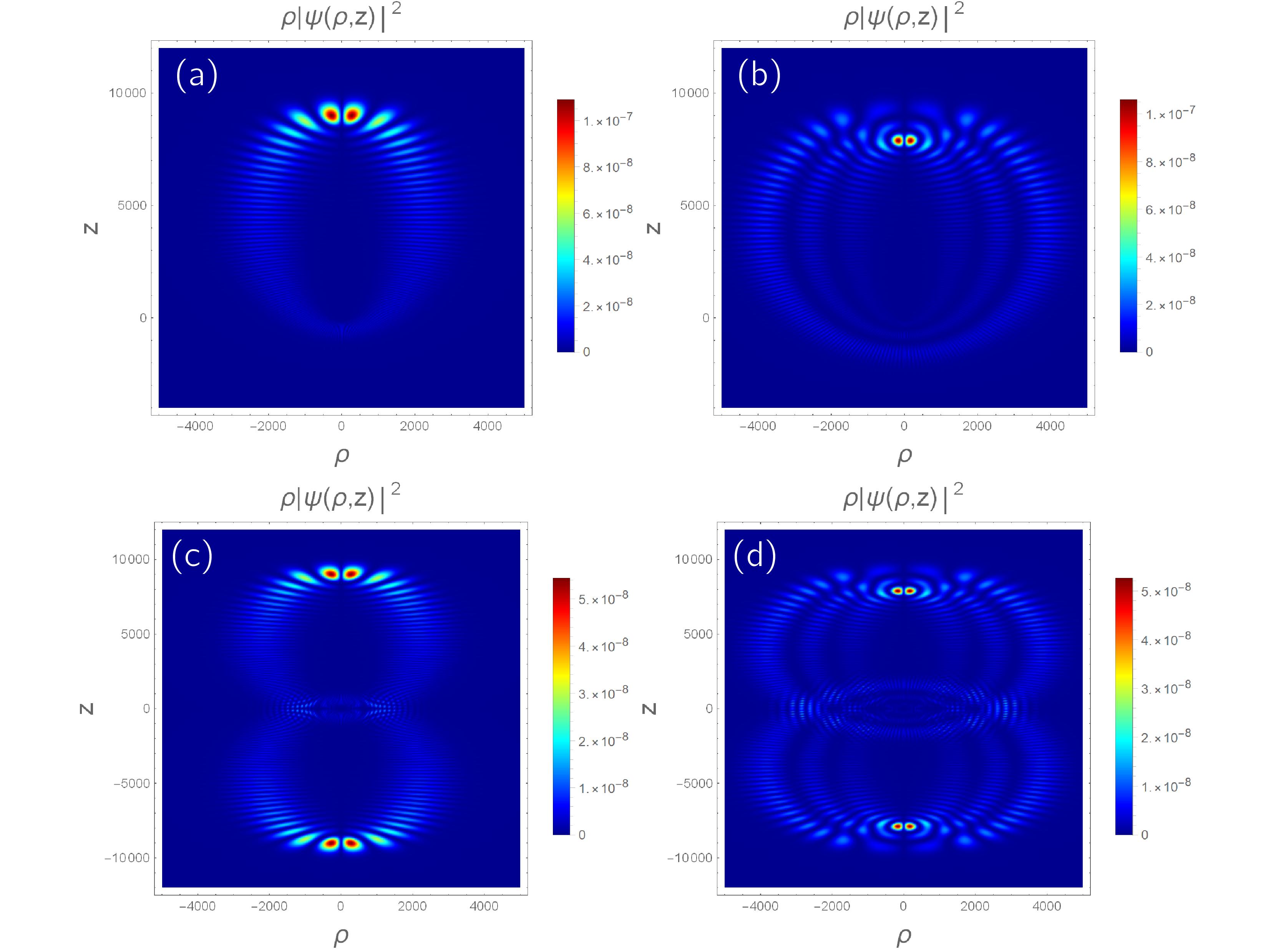}\\
\includegraphics[scale = 0.48]{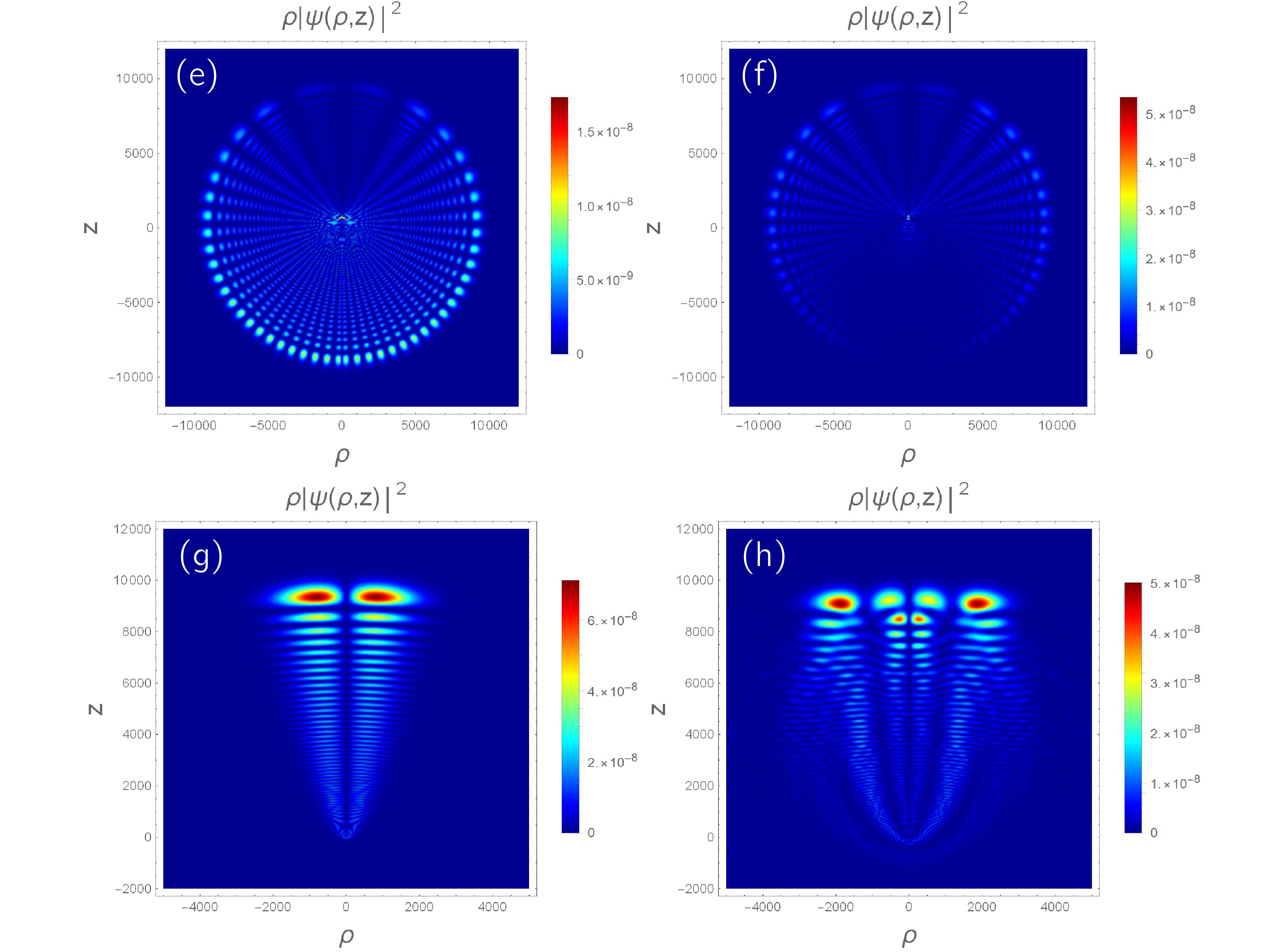}
\caption{\label{Fig:densitplots2}  Position-space Rydberg molecule wave functions, plotted in cylindrical coordinates.  a) $b=1$ trilobite; b) $b=3$ trilobite; c) $b=1$ trilobite trimer; d) $b=3$ trilobite trimer; e) $b=48$ trilobite; f) butterfly; g) stark state; h) $b=1$ trilobite, 84\% match. }
\end{figure}
\begin{figure}
\includegraphics[scale = 0.48]{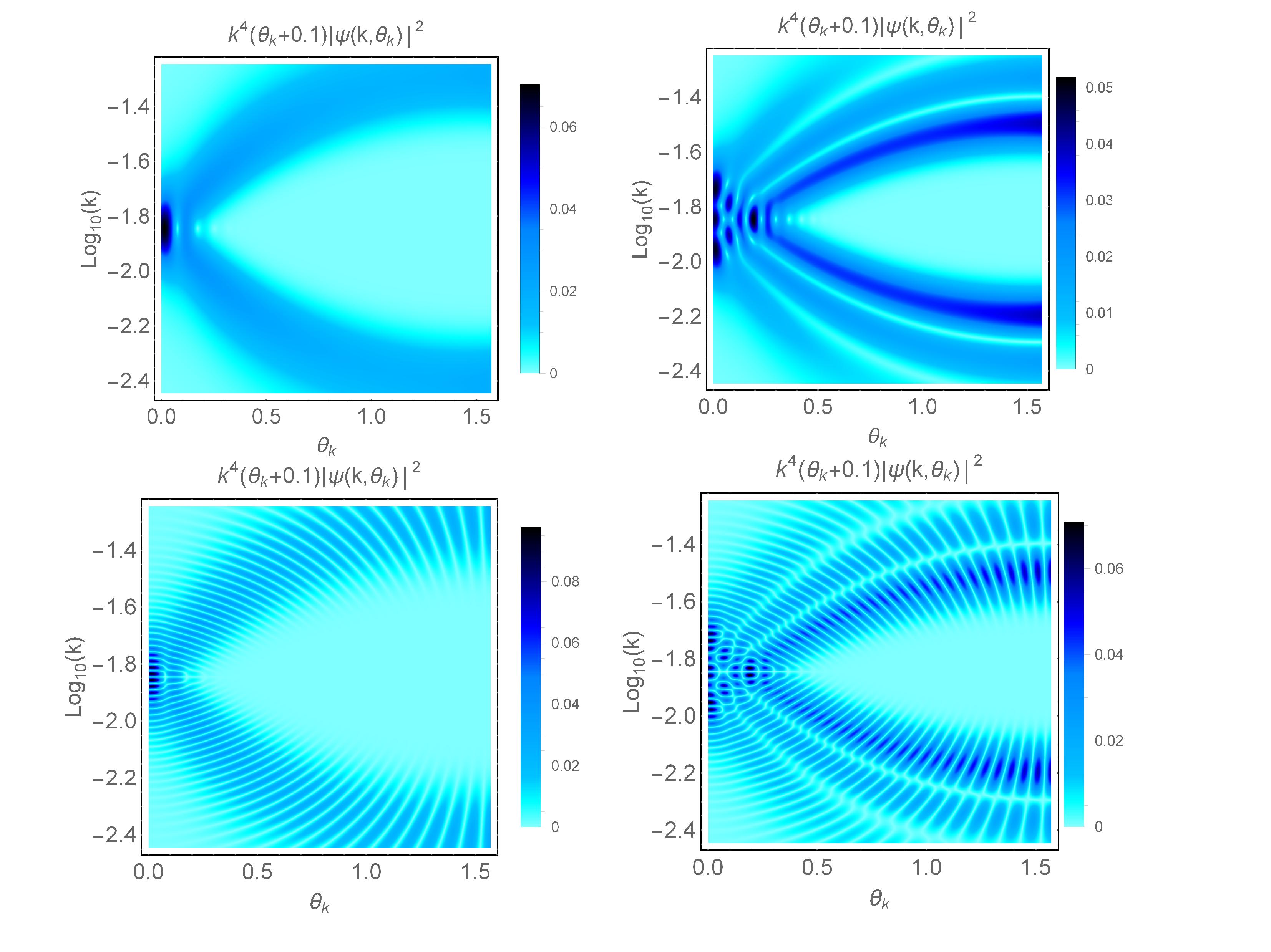}\\
\includegraphics[scale = 0.48]{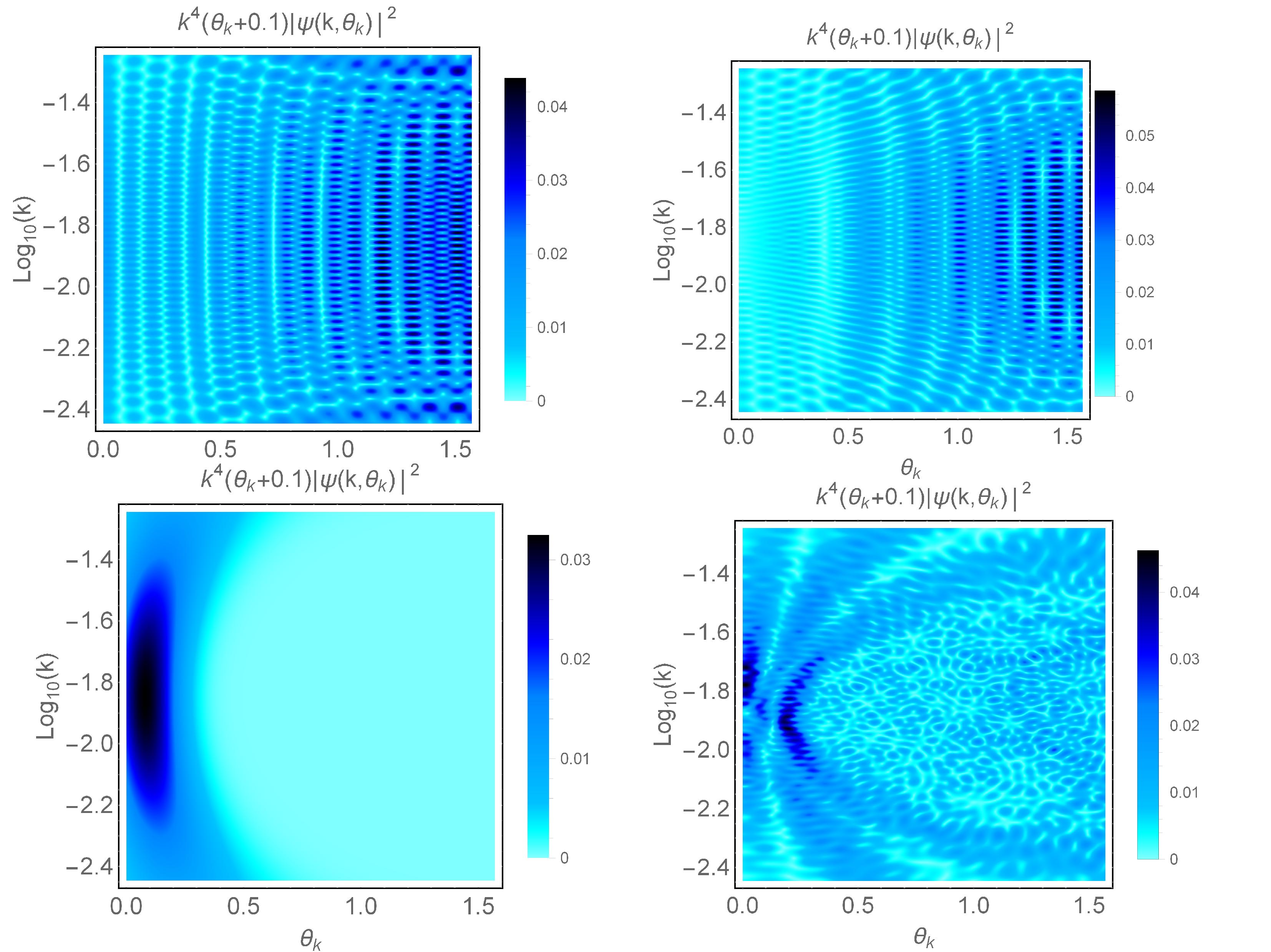}
\caption{\label{Fig:densitplotsmom1} Momentum-space Rydberg molecule wave functions, plotted in spherical coordinates. a) $b=1$ trilobite; b) $b=3$ trilobite; c) $b=1$ trilobite trimer; d) $b=3$ trilobite trimer; e) $b=48$ trilobite; f) butterfly; g) stark state; h) $b=1$ trilobite, 84\% match. }
\end{figure}
\begin{figure}[h]
\includegraphics[scale = 0.4]{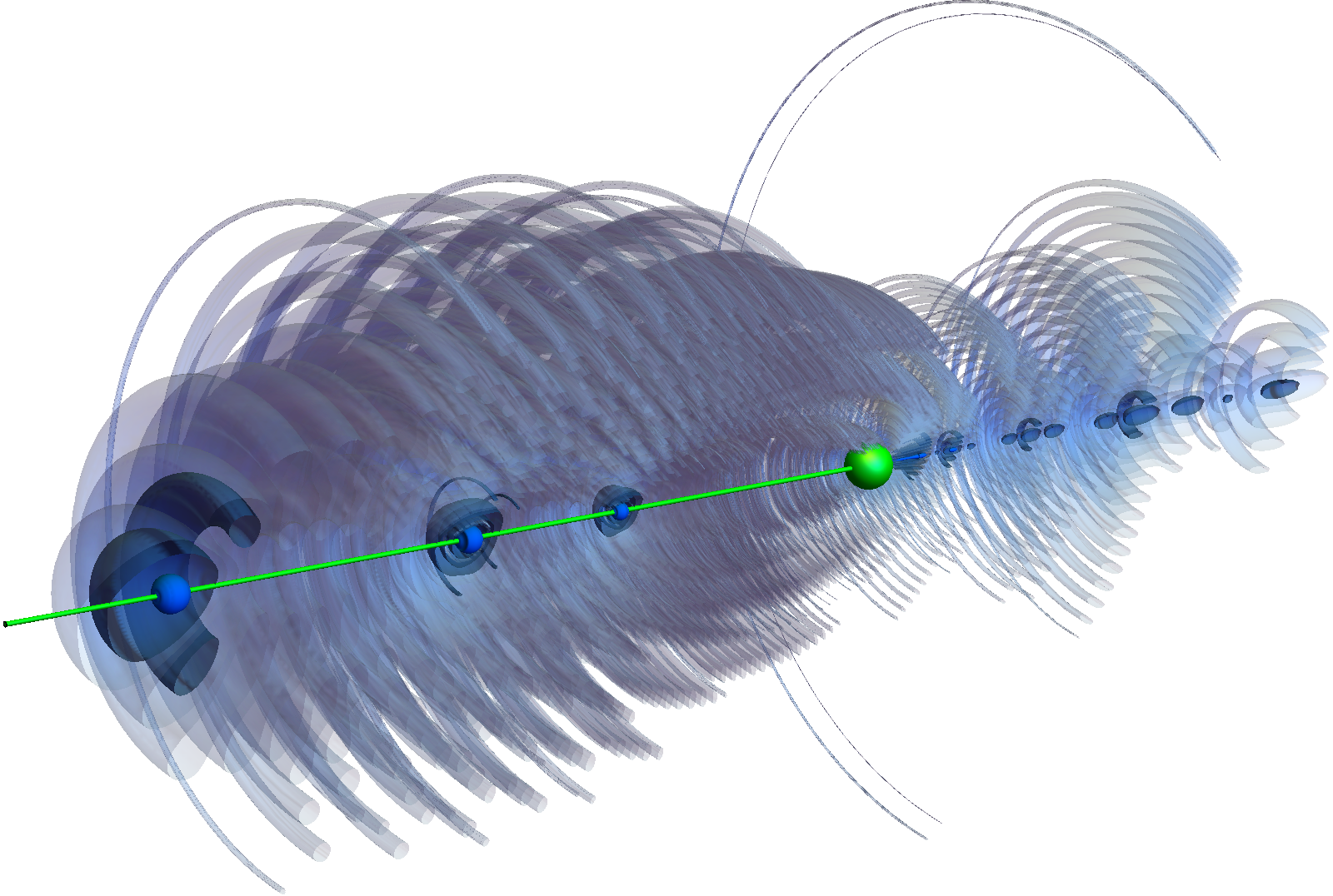}
\caption{\label{Fig:asd} Even more unusual variants can be formed following this same scheme, such as this case with three electron peaks placed along an axis. The color scheme is the same as in the main text, with the small bright blue regions having 10X the amplitude as the darker blue, and 100X the amplitude as the prominent gray regions. 
}
\end{figure}

\section{Data tables}
130-pulse sequences to make 99.9\% molecular bonds such as those shown in the first two figures in the main text.
\begin{center}
\begin{table}
\resizebox{450pt}{!}{%
\begin{tabular}{cc|cc||cc|cc||cc|cc||cc|cc}
$k=1$ & (1-65) & (66-130) & 143649 &$k=3$ & (1-65) & (66-130) & 144631 &trimer & (1-65) & (66-130) & 144583 &btfly & (1-65) & (66-130) & 144578 \\
\hline
 36.0129 & 239.973 & 32.2792 & 224.897 & 30.6755 & 285.278 & 28.2444 & 295.56 & 35.4941 & 317.394 & 34.9943 & 371.849 & 60.3706 & 367.018 & 60.7922 & 228.261 \\
 47.4666 & 296.119 & 34.6906 & 224.662 & 59.6629 & 244.097 & 58.1024 & 266.409 & 41.2067 & 227.01 & 61.6375 & 330.277 & 39.8348 & 200.876 & 58.1241 & 242.571 \\
 30.0045 & 371.947 & 52.899 & 399.063 & 22.901 & 227.233 & 32.7547 & 254.69 & 37.1052 & 359.677 & 24.865 & 352.076 & 26.4902 & 305.972 & 38.5931 & 399.63 \\
 26.9364 & 305.557 & 32.3173 & 369.321 & 43.4052 & 329.984 & 57.4415 & 252.275 & 22.7512 & 274.308 & 53.14 & 250.2 & 59.5579 & 220.389 & 45.8705 & 378.874 \\
 48.2474 & 390.364 & 21.6446 & 326.888 & 46.8757 & 205.56 & 57.1991 & 327.417 & 43.174 & 316.964 & 45.5835 & 343.93 & 35.2915 & 219.911 & 42.8275 & 286.915 \\
 35.3907 & 286.297 & 49.5366 & 256.09 & 52.8709 & 362.685 & 30.79 & 381.279 & 52.2115 & 320.772 & 31.6169 & 316.027 & 28.1597 & 206.653 & 36.5052 & 366.661 \\
 30.4024 & 323.885 & 39.0652 & 315.117 & 36.9635 & 298.669 & 26.7545 & 357.504 & 44.5726 & 303.069 & 30.7177 & 327.868 & 54.4036 & 259.798 & 36.5816 & 335.232 \\
 56.3999 & 209.329 & 44.7074 & 381.198 & 20.7992 & 395.597 & 23.3285 & 395.421 & 30.6065 & 352.557 & 50.4794 & 213.035 & 24.8793 & 296.894 & 25.1104 & 357.093 \\
 45.9804 & 236.636 & 21.7709 & 326.225 & 39.8652 & 274.949 & 46.741 & 383.931 & 30.5509 & 315.958 & 52.9686 & 292.449 & 43.5578 & 313.587 & 40.7838 & 328.844 \\
 21.4295 & 335.036 & 57.9015 & 352.016 & 29.2127 & 373.025 & 56.8444 & 210.272 & 54.4517 & 295.954 & 30.9039 & 355.396 & 56.7162 & 245.521 & 35.0073 & 236.966 \\
 22.8669 & 367.902 & 54.4536 & 290.009 & 48.6238 & 397.661 & 32.1329 & 370.697 & 58.3629 & 329.656 & 38.9009 & 376.617 & 37.5066 & 220.36 & 19.574 & 365.431 \\
 32.0231 & 383.247 & 31.783 & 255.278 & 58.3326 & 345.844 & 21.8744 & 356.343 & 50.6946 & 376.125 & 55.3135 & 335.137 & 20.4112 & 393.853 & 35.8386 & 321.852 \\
 60.2675 & 265.296 & 48.1839 & 315.093 & 45.4129 & 202.377 & 47.0193 & 276.46 & 38.8064 & 228.696 & 49.6757 & 363.387 & 40.9484 & 363.069 & 45.881 & 268.174 \\
 38.8887 & 223.329 & 60.4721 & 250.07 & 51.7152 & 395.18 & 46.6587 & 288.208 & 32.8958 & 319.173 & 32.5454 & 251.306 & 51.8996 & 363.292 & 33.9069 & 373.105 \\
 58.0465 & 334.239 & 48.1055 & 212.31 & 22.0069 & 341.64 & 41.3305 & 210.498 & 50.8416 & 340.081 & 47.9297 & 280.926 & 39.7599 & 249.473 & 33.4779 & 207.189 \\
 57.5267 & 389.306 & 35.8335 & 217.65 & 26.2229 & 242.293 & 41.0315 & 299.244 & 49.1342 & 238.827 & 34.8307 & 273.306 & 57.2002 & 399.043 & 34.7687 & 222.04 \\
 23.5209 & 394.01 & 38.2097 & 201.5 & 50.2489 & 299.515 & 23.5623 & 219.671 & 30.9119 & 200.408 & 44.5811 & 278.972 & 32.0786 & 282.291 & 29.7687 & 305.958 \\
 54.5002 & 295.453 & 29.6287 & 285.938 & 54.3202 & 286.47 & 43.2787 & 296.826 & 22.971 & 226.089 & 35.117 & 337.097 & 32.5831 & 322.687 & 37.8944 & 305.699 \\
 39.4703 & 206.331 & 32.3441 & 258.172 & 56.3983 & 307.395 & 24.8648 & 379.713 & 24.3993 & 219.168 & 43.812 & 236.388 & 20.0271 & 361.559 & 47.3182 & 360.582 \\
 56.7084 & 256.443 & 30.9668 & 263.233 & 55.9024 & 259.512 & 29.4882 & 202.415 & 52.3659 & 334.999 & 28.1357 & 246.272 & 27.731 & 272.172 & 34.3291 & 340.796 \\
 28.0783 & 230.737 & 46.2487 & 379.572 & 31.1048 & 200.4 & 53.7477 & 385.433 & 22.6289 & 341.545 & 58.1248 & 365.036 & 35.2022 & 338.487 & 34.9558 & 330.768 \\
 27.4159 & 360.588 & 54.245 & 337.983 & 39.6045 & 336.299 & 58.7858 & 351.59 & 36.6224 & 348.738 & 55.4131 & 282.814 & 40.0481 & 338.553 & 36.2144 & 356.28 \\
 28.0018 & 209.663 & 51.2241 & 254.757 & 26.7141 & 204.882 & 56.6222 & 247.432 & 50.4774 & 308.945 & 48.6759 & 227.929 & 30.2489 & 326.799 & 36.7335 & 264.212 \\
 41.1562 & 375.336 & 54.7605 & 249.41 & 41.1233 & 273.923 & 46.0954 & 381.257 & 28.8492 & 242.746 & 29.9056 & 268.698 & 24.4098 & 216.463 & 37.3114 & 275.256 \\
 23.6961 & 269.633 & 45.594 & 358.763 & 32.4852 & 281.958 & 35.3781 & 388.049 & 40.7027 & 235.288 & 42.5593 & 229.814 & 37.4246 & 291.68 & 55.8091 & 232.417 \\
 44.193 & 239.039 & 21.8992 & 259.108 & 38.1478 & 384.26 & 48.7241 & 396.525 & 26.8742 & 378.235 & 35.1144 & 352.779 & 33.0148 & 326.247 & 44.7961 & 248.977 \\
 46.8809 & 266.353 & 26.2047 & 314.306 & 32.0011 & 217.978 & 31.4193 & 296.609 & 47.0053 & 242.595 & 55.8567 & 205.326 & 21.356 & 256.536 & 22.7177 & 360.936 \\
 27.4893 & 233.389 & 60.5374 & 213.329 & 55.9752 & 288.14 & 51.5617 & 357.951 & 56.9714 & 333.463 & 55.9697 & 277.301 & 58.0937 & 307.07 & 36.4336 & 375.12 \\
 59.4849 & 241.986 & 22.2376 & 380.693 & 59.7606 & 348.423 & 26.5195 & 371.158 & 37.708 & 392.821 & 28.0426 & 252.5 & 35.9571 & 239.647 & 45.9658 & 304.406 \\
 22.6778 & 288.645 & 23.7563 & 237.295 & 51.8483 & 352.026 & 34.3525 & 300.214 & 53.9524 & 229.21 & 39.4129 & 295.992 & 49.3244 & 278.39 & 30.0952 & 379.697 \\
 31.5342 & 215.259 & 38.1811 & 364.905 & 24.84 & 351.879 & 54.7173 & 234.761 & 37.2985 & 219.555 & 37.558 & 383.275 & 34.4936 & 291.086 & 41.2862 & 314.257 \\
 47.0625 & 292.209 & 20.9276 & 246.949 & 58.3077 & 268.378 & 31.1896 & 369.958 & 53.0096 & 238.627 & 36.2796 & 220.377 & 50.1296 & 297.785 & 40.6829 & 232.539 \\
 36.8888 & 382.182 & 29.8878 & 378.93 & 44.0284 & 278.645 & 44.8019 & 269.558 & 26.9275 & 247.854 & 58.1631 & 326.799 & 23.9885 & 365.504 & 38.6288 & 225.26 \\
 21.9193 & 280.661 & 24.6508 & 282.762 & 41.7882 & 298.654 & 31.932 & 220.334 & 54.7396 & 250.558 & 33.0521 & 318.584 & 51.5307 & 296.675 & 22.5173 & 342.054 \\
 20.2117 & 368.671 & 34.157 & 345.272 & 32.4352 & 262.323 & 23.4262 & 334.959 & 31.6746 & 364.349 & 42.9386 & 392.974 & 53.9926 & 233.198 & 40.3963 & 241.13 \\
 34.0221 & 384.654 & 27.9355 & 250.431 & 45.3283 & 316.765 & 22.251 & 235.589 & 46.3985 & 339.204 & 45.9787 & 238.165 & 39.6537 & 279.727 & 25.4145 & 289.824 \\
 43.1697 & 352.925 & 39.9112 & 305.71 & 58.4723 & 232.633 & 25.6947 & 327.89 & 26.5834 & 342.254 & 46.7611 & 259.763 & 50.0239 & 353.974 & 47.5152 & 216.052 \\
 32.9205 & 270.138 & 53.0118 & 375.556 & 57.6169 & 314.096 & 26.5022 & 277.831 & 48.5323 & 282.514 & 26.0348 & 330.627 & 27.7551 & 222.77 & 42.6354 & 267.576 \\
 37.0737 & 212.815 & 39.453 & 368.708 & 46.1446 & 285.484 & 52.1491 & 257.077 & 52.9682 & 337.359 & 25.494 & 303.908 & 37.3639 & 374.937 & 49.3535 & 312.226 \\
 43.3936 & 367.695 & 37.5166 & 364.99 & 26.4938 & 306.158 & 54.108 & 305.788 & 48.049 & 398.049 & 45.5871 & 228.933 & 39.1577 & 220.815 & 25.2979 & 246.669 \\
 55.2011 & 366.178 & 26.5726 & 309.188 & 45.6729 & 399.616 & 58.3289 & 297.998 & 25.9886 & 242.361 & 36.7109 & 227.367 & 55.7195 & 354.727 & 33.6709 & 268.085 \\
 27.6309 & 260.188 & 45.2975 & 263.423 & 42.7266 & 348.977 & 32.0211 & 208.119 & 48.9908 & 325.304 & 47.0478 & 287.425 & 39.606 & 278.581 & 42.3664 & 274.643 \\
 42.9364 & 333.198 & 36.1652 & 214.373 & 35.4476 & 213.432 & 36.2285 & 260.191 & 30.7462 & 286.924 & 54.5021 & 350.246 & 40.0968 & 216.426 & 39.8083 & 295.178 \\
 40.0133 & 259.606 & 28.0009 & 273.357 & 30.2824 & 274.801 & 43.815 & 346.334 & 37.7796 & 360.876 & 30.659 & 258.053 & 46.6367 & 301.11 & 39.2659 & 273.952 \\
 33.0666 & 218.466 & 22.2618 & 341.86 & 40.8836 & 346.152 & 50.3911 & 387.497 & 59.8862 & 332.371 & 47.8343 & 344.693 & 46.8011 & 239.983 & 22.707 & 337.213 \\
 54.7871 & 373.83 & 33.7303 & 399.873 & 41.4858 & 328.858 & 34.361 & 327.299 & 25.0457 & 351.478 & 24.1661 & 205.396 & 23.2958 & 298.147 & 23.9924 & 398.095 \\
 54.3451 & 397.18 & 22.9229 & 224.574 & 53.4793 & 384.304 & 27.7848 & 291.853 & 24.7318 & 292.757 & 49.4903 & 319.222 & 49.2207 & 319.552 & 29.7442 & 319.683 \\
 57.8814 & 328.957 & 46.1466 & 327.839 & 56.6214 & 216.886 & 36.1084 & 210.678 & 37.0623 & 271.255 & 26.2631 & 213.913 & 31.7887 & 300.338 & 53.9314 & 372.497 \\
 31.5916 & 346.092 & 53.2207 & 213.49 & 40.233 & 264.668 & 38.8696 & 301.498 & 51.1656 & 246.016 & 39.0814 & 269.016 & 56.8078 & 283.341 & 45.7196 & 326.861 \\
 24.6633 & 231.851 & 46.8305 & 234.938 & 47.705 & 242.544 & 34.5773 & 389.282 & 50.2561 & 352.668 & 45.4389 & 345.164 & 43.4683 & 365.088 & 40.0147 & 381.293 \\
 54.3504 & 261.233 & 50.3357 & 344.638 & 29.0366 & 231.35 & 46.7124 & 363.397 & 36.3142 & 265.519 & 39.3281 & 396.8 & 40.7234 & 362.227 & 52.517 & 348.404 \\
 32.1983 & 389.8 & 36.9001 & 331.687 & 21.2776 & 223.807 & 32.3433 & 329.799 & 41.0068 & 275.675 & 26.0152 & 377.603 & 49.116 & 310.698 & 55.5193 & 380.661 \\
 22.6872 & 274.048 & 29.8008 & 215.523 & 28.3009 & 381.216 & 45.6109 & 352.837 & 55.9664 & 386.452 & 52.1098 & 221.565 & 33.4176 & 330.421 & 51.5434 & 322.372 \\
 43.4619 & 358.925 & 50.3533 & 207.864 & 19.7572 & 305.762 & 46.3676 & 378.444 & 29.6656 & 316.855 & 43.691 & 309.508 & 35.0741 & 336.073 & 25.6841 & 326.033 \\
 28.8527 & 252.449 & 52.4104 & 221.434 & 56.604 & 202.659 & 28.8499 & 233.439 & 34.2256 & 390.769 & 49.2657 & 312.438 & 27.3429 & 212.455 & 24.8583 & 251.831 \\
 39.4534 & 305.38 & 52.7095 & 205.81 & 21.9487 & 236.969 & 59.5223 & 366.226 & 38.1006 & 283.029 & 25.9424 & 387.424 & 44.9696 & 387.841 & 25.2758 & 266.513 \\
 30.43 & 241.009 & 21.9003 & 237.052 & 55.7136 & 306.774 & 22.2856 & 350.021 & 51.914 & 309.416 & 19.6119 & 290.035 & 37.1665 & 381.029 & 56.1465 & 274.791 \\
 27.3553 & 332.955 & 34.6828 & 332.034 & 31.3573 & 365.94 & 36.114 & 285.159 & 34.6322 & 371.574 & 53.2117 & 380.421 & 25.9724 & 392.491 & 29.341 & 206.092 \\
 54.0787 & 245.303 & 32.9304 & 269.907 & 35.8464 & 263.302 & 52.1255 & 285.945 & 32.4816 & 212.59 & 51.8562 & 359.448 & 27.8902 & 311.855 & 43.1052 & 279.411 \\
 37.2938 & 295.502 & 40.6183 & 344.606 & 23.9831 & 216.722 & 23.9845 & 353.351 & 35.1543 & 382.07 & 43.5743 & 204.431 & 22.9743 & 391.437 & 58.5682 & 369.634 \\
 56.3689 & 237.913 & 26.4251 & 212.98 & 52.1499 & 310.168 & 46.6141 & 238.038 & 43.5145 & 258.071 & 38.2452 & 374.988 & 40.0114 & 250.353 & 21.3546 & 301.212 \\
 40.6591 & 359.133 & 57.6148 & 379.082 & 59.4687 & 292.52 & 41.9842 & 396.399 & 33.234 & 209.186 & 39.4088 & 262.242 & 59.7528 & 316.438 & 42.0565 & 381.647 \\
 34.8486 & 322.082 & 25.6399 & 396.547 & 43.7048 & 375.261 & 26.1372 & 330.203 & 43.1673 & 337.063 & 22.405 & 350.151 & 20.0563 & 394.524 & 22.1436 & 311.549 \\
 52.8148 & 250.462 & 31.7913 & 205.877 & 35.9426 & 363.576 & 39.6143 & 308.127 & 29.5018 & 340.9 & 49.5556 & 332.431 & 22.7972 & 352.624 & 41.8479 & 250.578 \\
 55.5378 & 314.747 & 49.2313 & 381.603 & 59.3795 & 210.514 & 22.9112 & 270.785 & 29.643 & 373.777 & 57.0129 & 283.93 & 25.3702 & 383.803 & 59.5442 & 203.046 
\end{tabular}}
\caption{Pulse parameters. Top row: the type of bond (trilobite, $k=1,k=3$; triatomic trilobite with $k=1$; butterfly). Total time, including field ramps (both the up and down ramps last for 50$\mu$s), in nanoseconds. The left column of each is the duration of the electric field pulse, while the right column is the interval in between electric field pulses. 130 pulses are used each time. Times are always in ns. }
\end{table}
\end{center}

\end{document}